\documentclass{emulateapj}
\shorttitle{Resolving the Bright HCN(1 -- 0) Emission toward the
	Sy 2 Nucleus of M51}
\shortauthors{Matsushita et al.}

\slugcomment{ApJ, accepted}

\begin{document}

\title{Resolving the Bright HCN(1-0) Emission toward the Seyfert 2
	Nucleus of M51:\\
	Shock Enhancement by Radio Jets and Weak Masing by Infrared Pumping?}

\author{Satoki Matsushita\altaffilmark{1},
	Dinh-V-Trung\altaffilmark{2},
    Fr\'ed\'eric Boone\altaffilmark{3,4},
	Melanie Krips\altaffilmark{5},
	Jeremy Lim\altaffilmark{6},
	Sebastien Muller\altaffilmark{7}
	}

\altaffiltext{1}{Academia Sinica Institute of Astronomy and
	Astrophysics, P.O.\ Box 23-141, Taipei 10617, Taiwan, R.O.C.}
\altaffiltext{2}{Institute of Physics, Vietnamese Academy of Science
	\& Technology, 10, Daotan, BaDinh, Hanoi, Vietnam}
\altaffiltext{3}{Universit\'e de Toulouse, UPS-OMP, IRAP, Toulouse,
	France}
\altaffiltext{4}{CNRS, IRAP, 9 Av.\ colonel Roche, BP 44346, 31028,
	Toulouse Cedex 4, France}
\altaffiltext{5}{Institute de Radio Astronomie Millim\'etrique,
	300 Rue de la Piscine, 38406 Saint Martin d'H\`eres, France}
\altaffiltext{6}{Department of Physics, University of Hong Kong,
	Pokfulam Road, Hong Kong}
\altaffiltext{7}{Department of Earth and Space Sciences,
	Chalmers University of Technology, Onsala Space Observatory,
	SE-43992 Onsala, Sweden}

\begin{abstract}
We present high angular resolution observations of the HCN(1 -- 0)
emission (at $\sim1''$ or $\sim34$ pc), together with CO J = 1 -- 0,
2 -- 1, and 3 -- 2 observations, toward the Seyfert 2 nucleus of M51
(NGC 5194).
The overall HCN(1 -- 0) distribution and kinematics are very similar
to that of the CO lines, which have been indicated as the
jet-entrained molecular gas in our past observations.
In addition, high HCN(1 -- 0)/CO(1 -- 0) brightness temperature ratio
of about unity is observed along the jets, similar to that
observed at the shocked molecular gas in our Galaxy.
These results strongly indicate that both diffuse and dense gases are
entrained by the jets and outflowing from the AGN.
The channel map of HCN(1 -- 0) at the systemic velocity shows a
strong emission right at the nucleus, where no obvious emission has
been detected in the CO lines.
The HCN(1 -- 0)/CO(1 -- 0) brightness temperature ratio at this
region reaches $>2$, a value that cannot be explained considering
standard physical/chemical conditions.
Based on our calculations, we suggest infrared pumping and possibly
weak HCN masing, but still requiring an enhanced HCN abundance for
the cause of this high ratio.
This suggests the presence of a compact dense obscuring molecular gas
in front of the nucleus of M51, which remains unresolved at our
$\sim1''$ ($\sim34$ pc) resolution, and consistent with the Seyfert 2
classification picture.
\end{abstract}


\section{Introduction}
\label{intro}

Active Galactic Nuclei (AGNs) are generally believed to consist of
supermassive black holes (SMBHs) surrounded by dense material (i.e.,
molecular gas and dust) in a disk or a torus configuration
\citep[e.g.,][]{ant93}.
At spatial scales of $>100$ pc, a number of observations toward
nearby AGN host galaxies in molecular gas have uncovered strong
concentrations of dense molecular gas at the location of AGNs
\citep[e.g.,][]{koh01,koh08a}, and these nuclear components often
show smooth velocity gradients, suggestive of rotating disks or
tori around AGNs \citep[e.g.,][]{jac93,koh96,hsi08,hic09,san12}.

At finer spatial scales of $\sim10$ pc, however, molecular gas
observations toward some of the previously observed AGNs revealed
very different images from the past results observed at larger
spatial scales.
The nuclear molecular gas shows very disturbed features, such as
jet-entrained or outflowing motions from AGNs
\citep{mat07,kri11,com13} or infalling motions toward AGNs
\citep{dav09,mul09}, with little evidence of obscuring rotating
disks or tori at this spatial scale.
Where is the obscuring material?
In this paper, we target one of the nearest Seyfert galaxies, M51
(NGC 5194), to tackle these questions.

M51 hosts a Seyfert 2 nucleus \citep{ho97} with a pair of radio jets
\citep{for85,cra92}.
It is located at a distance of 7.1 Mpc \citep{tak06}, which is about
one half that of NGC 1068 \citep[14.4 Mpc;][]{tul88}, and therefore a
very suitable target to observe nuclear molecular gas that requires
very high {\it linear} resolution.
The molecular gas around the nucleus of M51 is dense
\citep[$>10^{4}$ cm$^{-3}$;][]{koh96} and warm
\citep[$>100$ K;][]{mat98,mat04}.
The molecular gas distribution and kinematics within $\sim1''$
($\sim34$ pc) from the AGN are complex, showing no clear evidence for
a disk or torus, but instead two separated features located at the
eastern and western sides of the nucleus \citep{mat07}.
The western clump shows a velocity gradient along the radio jets,
and its value matches well with that of the ionized gas along the
jets \citep{bra04}, strongly indicating that the western clump is
entrained by the jets and outflowing from the nucleus \citep{mat07}.

These previous $\sim1''$ resolution studies, however, have only been
done with CO lines, which might trace different gas components,
including diffuse molecular gas, and may confuse our view of the gas
distribution and kinematics close to the nucleus.
On the other hand, previous $\sim4''$ resolution HCN(1 -- 0) image
\citep{koh96,koh08b} is not high resolution enough to resolve the
detailed structures that have been resolved with the $\sim1''$ CO
observations.
In addition, \citet{aal12} reported an HCN(1 -- 0) emission
enhancement in the molecular outflows of Mrk 231 at a scale of about
1 kpc, suggesting that the HCN emission can be a good tool for
tracing outflows.
Observations toward the nucleus of M51 using the HCN(1 -- 0) line at
$1''$ or better resolution are therefore essential to understand the
physical and chemical interactions between molecular gas and AGN
jets.
Here, we present $\sim1''$ resolution HCN(1 -- 0), HCO$^{+}$(1 -- 0),
and HNC(1 -- 0) data toward the nuclear region of M51, together with
the $\lesssim1''$ resolution CO (1 -- 0), (2 -- 1), and (3 -- 2)
data, and compare the distributions and kinematics of these lines.
We then discuss the possible emission mechanism of the HCN line from
the nuclear region of M51.

\begin{figure}
\plotone{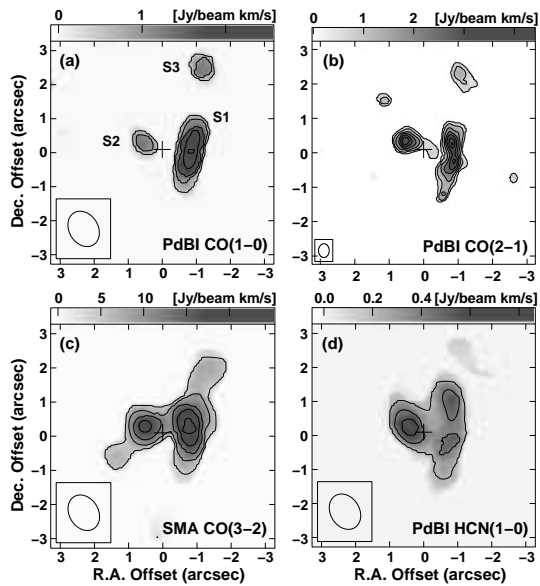}
\caption{Integrated intensity maps of (a) CO(1 -- 0), (b) CO(2 -- 1),
	(c) CO(3 -- 2), and (d) HCN(1 -- 0) lines toward the central
	$\pm3''$ ($\pm100$ pc) of M51 (the CO J = 2 -- 1 image is from
	\citealt{mat07}).
	The $uv$ coverages and the beam sizes ($1\farcs10\times0\farcs85$,
	P.A.\ $=31\arcdeg$) are matched, except the CO(2 -- 1) line map
	($0\farcs40\times0\farcs31$, P.A.\ $=0\arcdeg$).
	The central cross in each map displays the location of the type 2
	Seyfert nucleus determined from the peak of the 8.4~GHz radio
	continuum peak position of R.A.~$=13^{\rm h}29^{\rm m}52\fs7101$
	and Dec.~$=47\degr11\arcmin42\farcs696$ \citep{hag01,bra04}.
	In (a), three molecular gas clumps has been labeled with the same
	manner as in \citet{sco98}.
	Contour levels are as follows:
	(a) 2, 4, 6, 8, and 10 $\times$ 0.248 Jy beam$^{-1}$ km s$^{-1}$.
	(b) 1, 3, 5, 7, 9, and 11 $\times$
		0.334 Jy beam$^{-1}$ km s$^{-1}$.
	(c) 2, 4, 6, 8, and 10 $\times$ 2.21 Jy beam$^{-1}$ km s$^{-1}$.
	(d) 2, 4, and 6 $\times$ 0.109 Jy beam$^{-1}$ km s$^{-1}$.
\label{fig-mom0}}
\end{figure}

\section{Observation and Data Reduction}
\label{sect-obs}

\subsection{PdBI Observations}
\label{sect-obs-pdbi}

We observed the HCN(1 -- 0), HCO$^{+}$(1 -- 0), and HNC(1 -- 0) lines
simultaneously toward the nuclear region of M51 using the IRAM
Plateau de Bure Interferometer (PdBI).
The array was in the A configuration, and the observation was carried
out on February 1st, 2009.
The system temperatures in SSB were in the range 80 -- 400 K,
depending mostly on the elevation angle (i.e., atmospheric
absorption).
Three 160 MHz ($\sim535$ km s$^{-1}$) bandwidth correlators were
configured to cover each line.
Five 320 MHz units were configured primarily to cover nearby
line-free frequencies for continuum observations and calibrations.
\object{MWC 349} was observed for the flux calibration, the strong
quasar \object{3C273} was used for the bandpass calibration, and the
quasars \object{J1310+323} and \object{J1419+543} were used for the
phase and amplitude calibrations.
The uncertainty in the absolute flux scale is estimated to be around
5\%.

The data were calibrated using GILDAS, and were imaged using AIPS.
We did not detect any significant emission for the HCO$^{+}$(1 -- 0)
and HNC(1 -- 0) lines with the $3\sigma$ upper limit of
2.0 mJy beam$^{-1}$ in each 20 km s$^{-1}$ channel for both lines
within the velocity range of 350 -- 550 km s$^{-1}$.
We also did not detect any continuum emission at 90~GHz with the
$3\sigma$ upper limit of 0.87~mJy~beam$^{-1}$.

\begin{figure}
\plotone{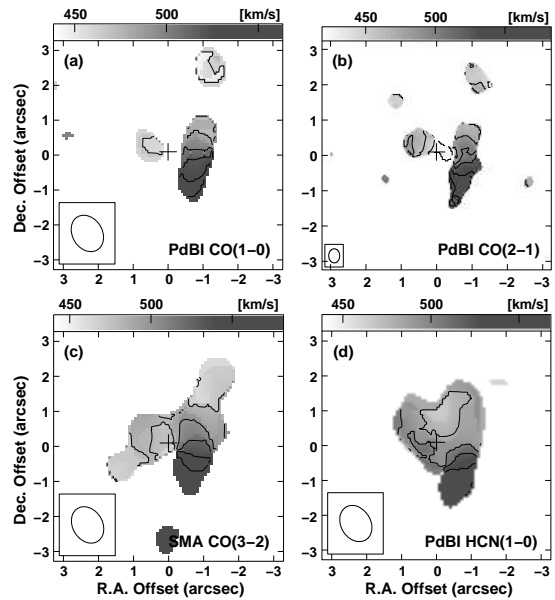}
\caption{Intensity-weighted velocity field maps of (a) CO(1 -- 0),
	(b) CO(2 -- 1), (c) CO(3 -- 2), and (d) HCN(1 -- 0) lines toward
	the central $\pm3''$ of M51 (the CO J = 2 -- 1 image is from
	\citealt{mat07}).
	Contour levels are as follows:
	(a) 460, 480, 500, ..., 560 km s$^{-1}$.
	(b) 440, 460, 480, ..., 540 km s$^{-1}$.
	(c) 460, 480, 500, ..., 560 km s$^{-1}$.
	(d) 480, 500, 520, 540, and 560 km s$^{-1}$.
	Other information is the same as in Fig.~\ref{fig-mom0}.
\label{fig-mom1}}
\end{figure}

\subsection{SMA Observations}
\label{sect-obs-sma}

The CO(3 -- 2) line was observed with the Submillimeter Array
\citep[SMA;][]{ho04}\footnotemark[1].
\footnotetext[1]{The Submillimeter Array is a joint project between
the Smithsonian Astrophysical Observatory and the Academia Sinica
Institute of Astronomy and Astrophysics, and is funded by
the Smithsonian Institution and the Academia Sinica.}
The observation was performed on April 18, 2005, and six out of eight
6~m antennas were used in the extended configuration.
The correlator was configured to a 2~GHz bandwidth with a 0.8125 MHz
channel resolution, and the line was located at the upper side band
(USB).
Titan was observed as a flux calibrator, and Uranus, Ganymede, 3C279,
and 3C345 were used as bandpass calibrators.
J1153+495 and 3C345 were observed every $\sim20$ minutes for the phase
and amplitude calibrations.
The uncertainty in the absolute flux scale is estimated to be around
15\%.

\begin{figure*}
\plotone{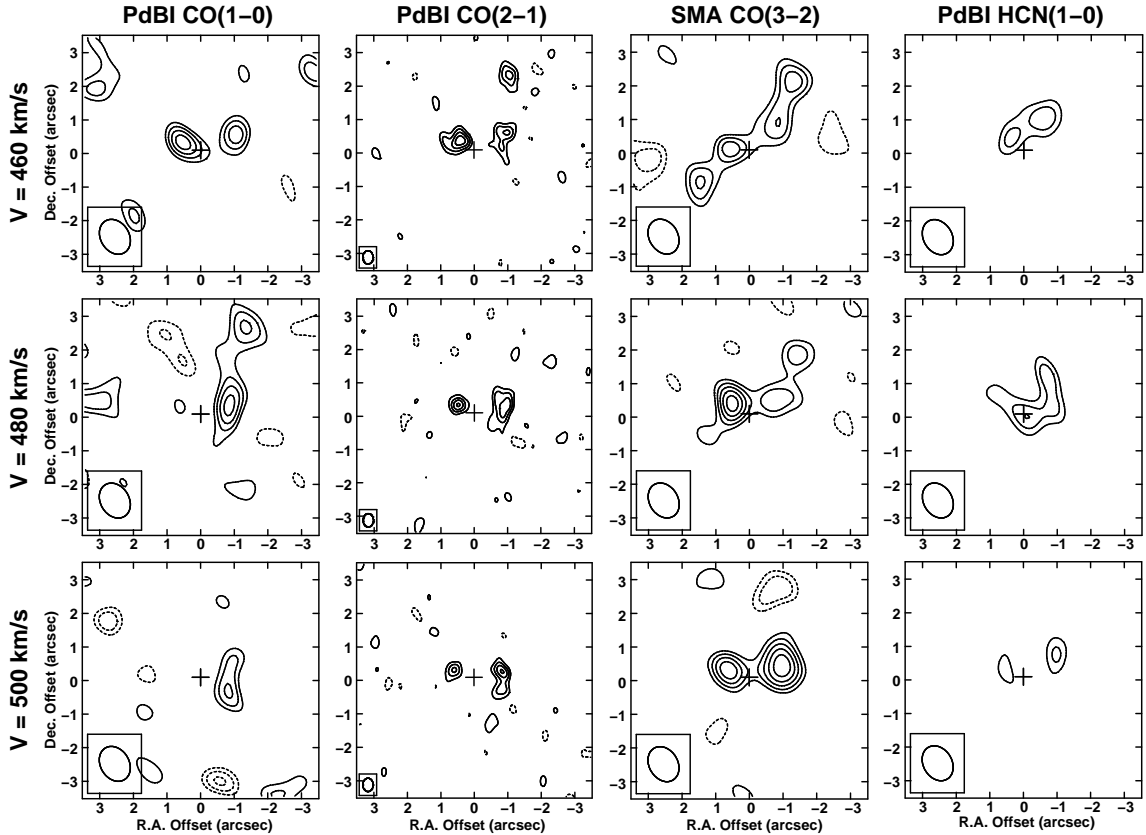}
\caption{Central $\pm3''$ channel maps between the velocity range of
	$480\pm20$ km s$^{-1}$ (i.e., systemic velocity channel map and
	one channel before and after).
	First to fourth columns are the CO(1 -- 0), CO(2 -- 1),
	CO(3 -- 2), and HCN(1 -- 0) lines, respectively.
	First, second, and third rows are the channel maps at the
	velocities of 460, 480, and 500 km s$^{-1}$, respectively.
	Contour levels are as follows:
	CO(1 -- 0): $-4$, $-3$, 3, 4, 5, and 6 $\times1\sigma$
		(= 4.6 mJy beam$^{-1}$ = 0.45 K).
	CO(2 -- 1): $-3$, 3, 5, 7, and 9 $\times1\sigma$
		(= 5.2 mJy beam$^{-1}$ = 0.96 K).
	CO(3 -- 2): $-3$, 3, 4, 5, 6, and 7 $\times1\sigma$
		(= 39.7 mJy beam$^{-1}$ = 0.43 K).
	HCN(1 -- 0): $-3$, 3, 4, and 5 $\times1\sigma$
		(= 1.9 mJy beam$^{-1}$ = 0.31 K).
	Note that there was no positive or negative noise more than
	$3\sigma$ in the HCN(1 -- 0) map.
	See Fig.~\ref{fig-mom0} caption for other information.
\label{fig-ch480kms}}
\end{figure*}

The data calibrations were done using the software package MIR,
adapted for SMA\footnotemark[2].
\footnotetext[2]{\url{http://www.cfa.harvard.edu/$\sim$cqi/mircook.html}}
This extended configuration dataset was combined with the previous
compact configuration data set \citep{mat04} in the $uv$ plane.
The combined data were CLEANed and the images were produced using
AIPS.

\subsection{Matching $uv$ Ranges}
\label{sect-obs-uv}

In this paper, we compare the newly obtained HCN(1 -- 0) and
CO(3 -- 2) data with the previously obtained PdBI A configuration
CO(1 -- 0) data \citep{mat07} in various ways, including line ratios.
For this purpose, we need to match the shortest $uv$ lengths and the
beam sizes of all the datasets, so that they all sample the same
structures (i.e., the same spatial frequency components obtained by
interferometers).
Note that matching the beam size has the same effect as matching the
longest $uv$ lengths.
The shortest $uv$ length common to all datasets is imposed by the
CO(1 -- 0) data, for which the shortest $uv$ length is 37 k$\lambda$
(i.e., we can image the structure smaller than $\sim6''$).
We then convolved all the images to the largest beam size among our
data of $1\farcs10\times0\farcs85$ with the position angle (P.A.) of
$31\arcdeg$.
We note here that this $uv$ range matching does not significantly
change the images from the $uv$ unlimited ones.
We did not match the $uv$ range for the CO(2 -- 1) data, since there
is a small overlap in the $uv$ range with other data (i.e., CO
J = 2 -- 1 data has the shortest $uv$ length twice longer than that
of J = 1 -- 0 data, since those data are taken simultaneously with
PdBI).
As we will show, the CO(2 -- 1) line traces the same features, and is
shown for reference, although it is not included in the line ratio
calculations.

\section{Results}
\label{sect-res}

In Fig.~\ref{fig-mom0}, we show the central $\pm3''$ ($\pm100$ pc)
region integrated intensity maps of the CO(1 -- 0), CO(2 -- 1),
CO(3 -- 2), and HCN(1 -- 0) lines.
All the data except CO(2 -- 1) are $uv$ and beam matched images as
mentioned above.
The spatial resolution in CO(2 -- 1) is more than a factor of two
higher than other images \citep{mat07}, and is shown here for
reference.
All the images appear to have similar overall molecular gas
distributions and kinematics:
There are two clumps located at the western and eastern sides of the
nucleus (hereafter clumps S1 and S2, respectively; see
Fig.~\ref{fig-mom0}a), whereby the western clumps S1 are elongated
along the north-south direction.
The intensity-weighted velocity maps for all the lines are shown in
Fig.~\ref{fig-mom1}.
In all the maps, the eastern clump S2 shows increasingly higher
blueshifted velocities toward the nucleus, and the western clump S1
shows clear velocity gradient in the north-south direction, as was
previously reported by \citet{mat07}.
The similarity in the kinematics for all the lines suggests that the
overall molecular gas distribution and kinematics at the central
$\pm3''$ does not depend on the molecular species (both CO and HCN)
or properties (both diffuse and dense gas, and from lower-J to
higher-J lines).
As mentioned in Sect.~\ref{intro}, the S1 clump is the jet-entrained
molecular gas, so that the similarity of the distribution and
kinematics between the CO and HCN data strongly indicates that
{\it the both diffuse and dense molecular gas is entrained by the
jets and outflowing from the AGN.}

\begin{figure}
\plotone{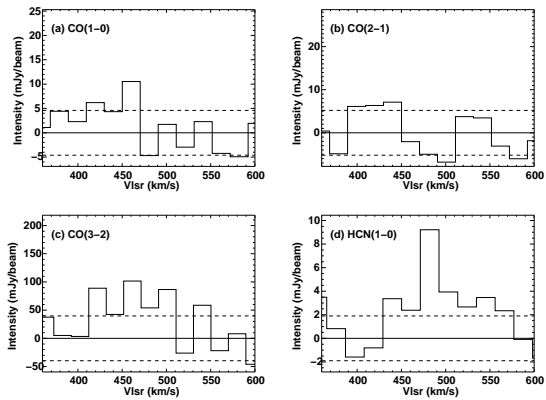}
\caption{Spectra toward the M51 nucleus for (a) CO(1 -- 0),
	(b) CO(2 -- 1), (c) CO(3 -- 2), and (d) HCN(1 -- 0) lines.
	The spectra have been taken at the position of the nucleus with
	the spatial resolution the same as the beam sizes.
	The nuclear position and the beam sizes are mentioned in the
	caption of Fig.~\ref{fig-mom0}.
	Horizontal dashed lines in each plot are $\pm1\sigma$ of each
	data set (see the caption of Fig.~\ref{fig-mom0} for the values).
	All the CO lines exhibit at most $\sim2\sigma$ peaks, but the
	HCN(1 -- 0) line only at the velocity channel of 480 km s$^{-1}$,
	which includes the systemic velocity of $471.7\pm0.3$ km s$^{-1}$
	\citep{she07}, shows a $\sim5\sigma$ peak.
\label{fig-spec}}
\end{figure}

With a detailed inspection of the channel maps, however, a distinct
difference between the CO and HCN distributions can be seen in a
channel map at the systemic velocity
\citep[$471.7\pm0.3$ km s$^{-1}$][]{she07}.
Fig.~\ref{fig-ch480kms} displays three channel maps centered at the
velocity of 460, 480, and 500 km s$^{-1}$, each with a width of
$\sim20$ km s$^{-1}$.
In the channel maps at the velocity of 460 and 500 km s$^{-1}$
(i.e., adjacent to the channel at the systemic velocity), all the
lines show the emission peaks at either or both eastern and western
sides of the nucleus (some emissions elongate toward the nucleus, but
do not peak there).
On the other hand, in the channel maps at the velocity of
480 km s$^{-1}$, which includes the systemic velocity of
$471.7\pm0.3$ km s$^{-1}$, {\it the HCN(1 -- 0) line clearly
exhibits a $\sim5\sigma$ peak at the nucleus, where CO lines show no
peak if any detectable emission at the nucleus}.
To further confirm this systemic velocity HCN emission peak, we plot
the spectra toward the nucleus of M51 for all four lines in
Fig.~\ref{fig-spec}.
As can be seen, all the CO lines exhibit at best emission at the
$\sim2\sigma$ level, but the HCN(1 -- 0) line shows a $\sim5\sigma$
peak only at the systemic velocity channel of 480 km s$^{-1}$.
This therefore indicates that {\it the central HCN(1 -- 0) emission
is highly related to the molecular gas around the systemic velocity}.

\begin{figure}
\plotone{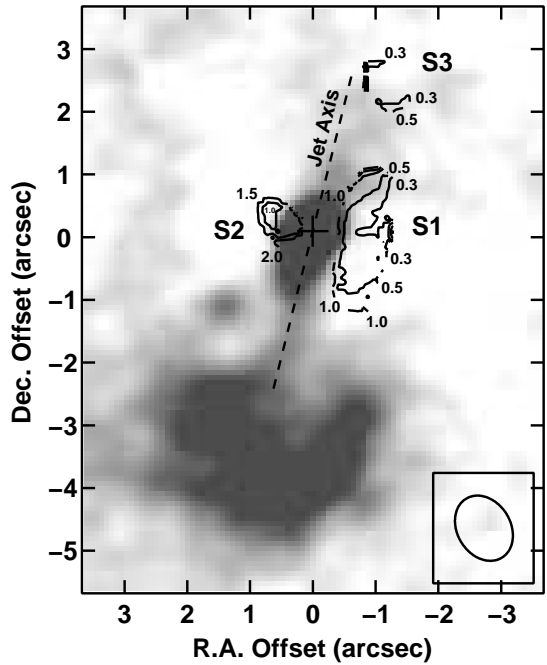}
\caption{HCN(1 -- 0)/CO(1 -- 0) brightness temperature ratio map
	(contours) overlaid on the VLA 6 cm radio continuum image
	\citep[greyscale;][]{cra92}.
	Contour levels are presented in the map.
	The dashed line shows the radio jet axis
	\citep[P.A.\ = $166^{\arcdeg}$;][]{bra04}, and the continuum
	image is enhanced to show the weak jet emission elongated toward
	south-east, and connected to the bow-shock emission at south-east
	of this image.
	The average ratios for the clumps S1, S2, and S3 are calculated
	as 0.6, 1.5, and 0.3, respectively.
	S1 is clearly elongated along this jet emission (see also
	\citealt{mat07}), and the higher ratio regions are concentrated
	where S1 and the jet overlap.
	S2 mostly overlaps with the jet emission and has high ratio
	overall, but higher toward the Seyfert 2 nucleus.
    Cross at the center is the same as in Fig.~\ref{fig-mom0}.
\label{fig-ratiomap}}
\end{figure}

After converting the integrated intensity maps of the CO(1 -- 0) and
HCN(1 -- 0) lines in Fig.~\ref{fig-mom0} into brightness temperature
maps, we made the HCN(1 -- 0)/CO(1 -- 0) brightness temperature ratio
map in Fig.~\ref{fig-ratiomap} overlaid on the VLA 6 cm continuum map
\citep{cra92}.
All the molecular gas clumps show the HCN/CO ratios of more than 0.1,
and the clumps closer in projection either to the AGN or to the jet
tend to have higher ratios.
Indeed, the highest ratio of more than 2.0 is located at the AGN
(western) side of the clump S2.
More interestingly, {\it all the eastern side of the clump S1 has
higher ratio of $\sim1.0$ than the western side of $\sim0.3-0.5$,
and this side is where the radio jets are running through}, which
can be clearly seen in Fig.~\ref{fig-ratiomap}.

The average ratios of the clumps S1, S2, and S3 are calculated as
0.6, 1.5, and 0.3, respectively.
On the other hand, since the CO(1 -- 0) emission totally lacks at the
nucleus (see Fig.~\ref{fig-mom0}a), the HCN/CO ratio at the nucleus
cannot be calculated in the ratio map.
But since the HCN(1 -- 0) emission does exist at the nucleus with the
integrated intensity of 0.422 Jy beam$^{-1}$ km s$^{-1}$, the lower
limit of the HCN/CO brightness temperature ratio at the nucleus can
be calculated as $>0.96$ using the $3\sigma$ upper limit of CO of
0.744 Jy beam$^{-1}$ km s$^{-1}$.
This is also true in the channel map at the velocity of
480 km s$^{-1}$; in Fig.~\ref{fig-ch480kms}, no clear emission in
CO(1 -- 0) toward the nucleus with $1\sigma$ noise level of 0.45 K,
while HCN(1 -- 0) is detected with 1.5 K ($\sim5\sigma$).

\begin{figure}
\plotone{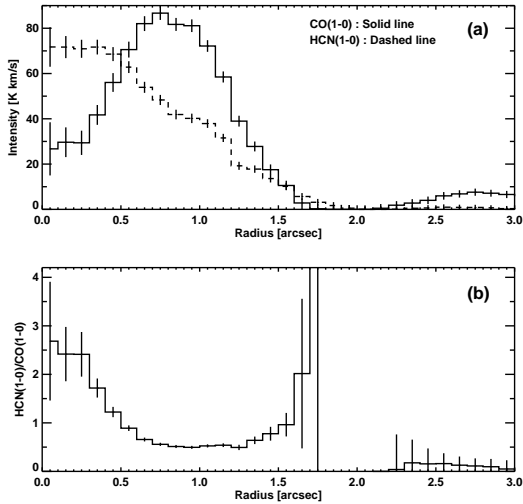}
\caption{(a) Central $3''$ radial intensity distributions of the
	CO(1 -- 0) (solid line) and HCN(1 -- 0) (dashed line) emission
	derived from the integrated intensity maps in
	Fig.~\ref{fig-mom0} after converted into brightness temperature
	scale.
	(b) Central $3''$ HCN/CO integrated intensity ratio distribution.
	The $uv$ coverages are matched between these two lines.
\label{fig-iring}}
\end{figure}

The radial tendency of the HCN/CO ratio is more clearly seen in the
radial ratio plot.
Fig.~\ref{fig-iring}(a) plots the radial integrated intensity
distributions of HCN(1 -- 0) and CO(1 -- 0) in unit of K km s$^{-1}$.
It is obvious that the HCN(1 -- 0) emission is centrally
concentrated, while the CO(1 -- 0) emission exhibits deficiency at
the nucleus.
From these two radial distributions, we made the radial
HCN(1 -- 0)/CO(1 -- 0) integrated intensity ratio plot in
Fig.~\ref{fig-iring}(b).
This figure clearly displays that the HCN/CO ratio is higher at the
nucleus and decreases monotonically as the distance from the nucleus
increases out to $\sim1\farcs25$.

As mentioned in Sect.~\ref{sect-obs-pdbi}, we did not detect the
HCO$^{+}$(1 -- 0) and HNC(1 -- 0) lines and therefore the maps for
these lines are not presented in this paper.
The brightness temperature ratios for both
HCN(1 -- 0)/HCO$^{+}$(1 -- 0) and HCN(1 -- 0)/HNC(1 -- 0) are
therefore $>4.3$ using $3\sigma$ upper limits for these non-detected
lines.

\section{Discussion}
\label{sect-dis}

\subsection{High HCN/CO Ratios at the Nucleus and Outflowing Gas}
\label{sect-dis-highratio}

Our high spatial resolution molecular gas images toward the Seyfert 2
nucleus of M51 clearly exhibit higher HCN(1 -- 0)/CO(1 -- 0) ratios
in the molecular gas clumps located close to the AGN than those in
other galaxies;
the jet-entrained molecular gas clump (clump S1; projected distance
of $\sim30$ pc from the nucleus) has an averaged HCN/CO ratio of 0.6,
and the nearest clump from the AGN (clump S2; projected
distance $<20$ pc) a ratio of 1.5.
The clump S3, which located farthest from the AGN, has a ratio of 0.3.

In normal and starburst galaxies, the HCN(1 -- 0)/CO(1 -- 0) ratios
are $<0.3$ \citep[e.g.,][]{sol92,hel93,aal95,sor02,gao04,koh08a,mat10}.
The ratios in our Galaxy and M31 are also low, around 0.01 -- 0.03
for Giant Molecular Clouds (GMCs) in the galactic disks
\citep{hel97,bro05}, and up to $\sim0.15$ at the Galactic center
\citep{jac96}.
The ratios can be up to $\sim0.6$ for the case of molecular gas
located around Seyfert nuclei \citep[e.g.,][]{koh08a,kri07,kri11}.
Based on these values, the averaged HCN/CO ratio of the clump S1 is
similar to those of other Seyfert galaxies, and that of the clump S3
is at the high end of normal or starburst galaxies.
The clump S2 is exceptional; the HCN intensity is brighter than that
of CO, and this has never been observed in other galaxies in the
past.

\subsection{Shocked Molecular Gas Outflow along the Radio Jets}
\label{sect-dis-shock}

As shown in Sect.~\ref{sect-res} and in Fig.~\ref{fig-ratiomap}, the
S1 clump show a high HCN/CO ratio of around unity only at the side
where the radio jets overlap.
In addition, this clump is known to be an outflow that is entrained
by the radio jets \citep{mat07}.
Such a high ratio has been observed at shocked regions of molecular
outflows from young stellar objects in our Galaxy.

\citet{ume92} observed a highly collimated outflow in the dark cloud
L1157 with the $^{12}$CO(1 -- 0), $^{13}$CO(1 -- 0), and HCN(1 -- 0)
lines.
In their $^{12}$CO spectrum, two (narrow and broad) components can be
seen, which are tracing quiescent and outflowing gas, respectively.
In the $^{13}$CO spectrum, only a narrow component has been detected,
but on the contrary, only a broad component has been detected in the
HCN spectrum.
Using their data, we derive the line ratios with assuming that the
HCN emission is only from the outflow component and the $^{13}$CO
emission from quiescent component.
For the $^{12}$CO line, we fit a Gaussian to each component to
derive the integrated intensity.
We integrate all the hyperfine components of the HCN line, since our
M51 data cannot distinguish these lines.
From these data, we calculated the HCN/$^{12}$CO and HCN/$^{13}$CO
ratios for the outflowing (shocked) gas as 0.9 and $>260$ (using
$3\sigma$ upper limit for the $^{13}$CO data), respectively.
Interferometric HCN and $^{12}$CO line observations of the Orion-KL
high velocity outflow \citep{wri96} also show high HCN/$^{12}$CO
ratios ranging between $0.4-0.5$.
The ratios in the Orion-KL outflow is lower than that in the L1157
outflow, possibly due to more complex nature (and therefore larger
contamination) in the massive star forming regions than in the low
mass star forming regions.

The ratio and the condition of the clump S1, especially the side
close to the radio jets, is very similar to that of the outflowing
(shocked) molecular gas in L1157.
This strongly supports the idea that the high HCN/CO ratio along the
radio jets in M51 is caused by the strong interaction (i.e., shock)
between the jets and the outflowing molecular gas.
Furthermore, it is worth mentioning that the averaged HCN/CO ratio of
0.6 is very similar to the dense molecular gas outflow
observed toward Mrk 231 of $\sim0.6$ \citep{aal12}.
Since their spatial resolution of $\sim1.2$ kpc is lower than ours,
it is possible that the actual HCN/CO ratio in Mrk 231 is higher.

\subsection{Infrared Pumping with Weak Maser at Circumnuclear
	Molecular Gas?}
\label{sect-dis-S2}

The clump S2 shows extremely high HCN/CO ratio of 1.5 on average, and
more than 2 at the nucleus as shown in Fig.~\ref{fig-iring}.
As mentioned above, such high ratios have not been observed in
extragalactic molecular gas, but observed in molecular gas around
Asymptotic Giant Branch (AGB) stars.

We observed the high mass-loss carbon AGB stars, IRC$+10216$, with
$^{12}$CO(1 -- 0), $^{13}$CO(1 -- 0), and HCN(1 -- 0) for the
intensity calibrations of our past observations
\citep[see][for details]{mat10}.
The resultant $^{12}$CO/$^{13}$CO, HCN/$^{12}$CO, and HCN/$^{13}$CO
integrated intensity ratios are $\sim11$, 1.3, and 15, respectively.
Note again that we integrate all the HCN hyperfine lines for this
calculation.
The high HCN/$^{12}$CO and HCN/$^{13}$CO ratios are due to the
infrared (IR) pumping \citep{din00}.

This result suggests that the extremely high HCN/CO ratio can be due
to the IR pumping.
In the following subsections, we discuss what is the possible
excitation mechanism for the clump S2.

\subsubsection{Excitation Mechanisms of the HCN Emission:
	I. LTE/non-LTE Conditions}
\label{sect-dis-lte}

We start by assuming that both molecules are excited solely by
collisions with H$_{2}$ molecules, which is typically the case in
Galactic GMCs.
The opacity ratio under the local thermal equilibrium (LTE) can be
expressed as
\begin{eqnarray}
\label{eq-tauratio}
\frac{\tau_{\rm HCN(1-0)}}{\tau_{\rm CO(1-0)}}
 & = & \frac{f_{\rm HCN}}{f_{\rm CO}}
       \left(\frac{\mu_{\rm HCN}}{\mu_{\rm CO}}\right)^{2}
       \frac{B_{\rm HCN}}{B_{\rm CO}}
       \frac{\rm [HCN]}{\rm [CO]}
       \nonumber \\
 &   & \times \frac{1-\exp(-h\nu_{\rm HCN(1-0)}/kT)}
                   {1-\exp(-h\nu_{\rm CO(1-0)}/kT)},
\end{eqnarray}
where $f$ is the filling factor, $\mu$ is the dipole moment, $B$ is
the rotational constant, $h$ is the Planck constant, $\nu$ is the
rotational transition frequency, and $k$ is the Boltzmann constant.
The HCN and CO molecular abundances relative to H$_{2}$ are expressed
as [HCN] and [CO], respectively.
Adopting $f_{\rm HCN}=f_{\rm CO}$, which is already an extreme case
compared to the Galactic GMCs where $f_{\rm HCN} \ll f_{\rm CO}$,
$\mu_{\rm HCN} = 2.984$, $\mu_{\rm CO} = 0.112$,
$B_{\rm HCN} = 44.316$ GHz, $B_{\rm CO} = 57.636$ GHz,
$\nu_{\rm HCN} = 88.631$ GHz, $\nu_{\rm CO} = 115.271$ GHz,
[HCN] $=2\times 10^{-8}$ \citep{irv87}, and [CO] $=5\times 10^{-5}$,
the opacity ratio can be calculated as
$\tau_{\rm HCN(1-0)}/\tau_{\rm CO(1-0)}\sim0.18\pm0.01$
under a large range of molecular gas temperature condition
($T\sim3-1000$ K).
The brightness temperature ratio can therefore be expressed as
\begin{eqnarray}
\label{eq-tempratio}
\frac{T_{\rm HCN(1-0)}}{T_{\rm CO(1-0)}}
 &  =   & \frac{1-\exp(-\tau_{\rm HCN})}{1-\exp(-\tau_{\rm CO})}
          \nonumber \\
 & \sim & \frac{1-\exp(-0.18\tau_{\rm CO})}
               {1-\exp(-\tau_{\rm CO})},
\end{eqnarray}
and this ratio ranges between 0.18 (optically thin limit) and 1
(optically thick limit), which cannot realize
HCN(1 -- 0)/CO(1 -- 0) $>1$.

\begin{figure*}
\plotone{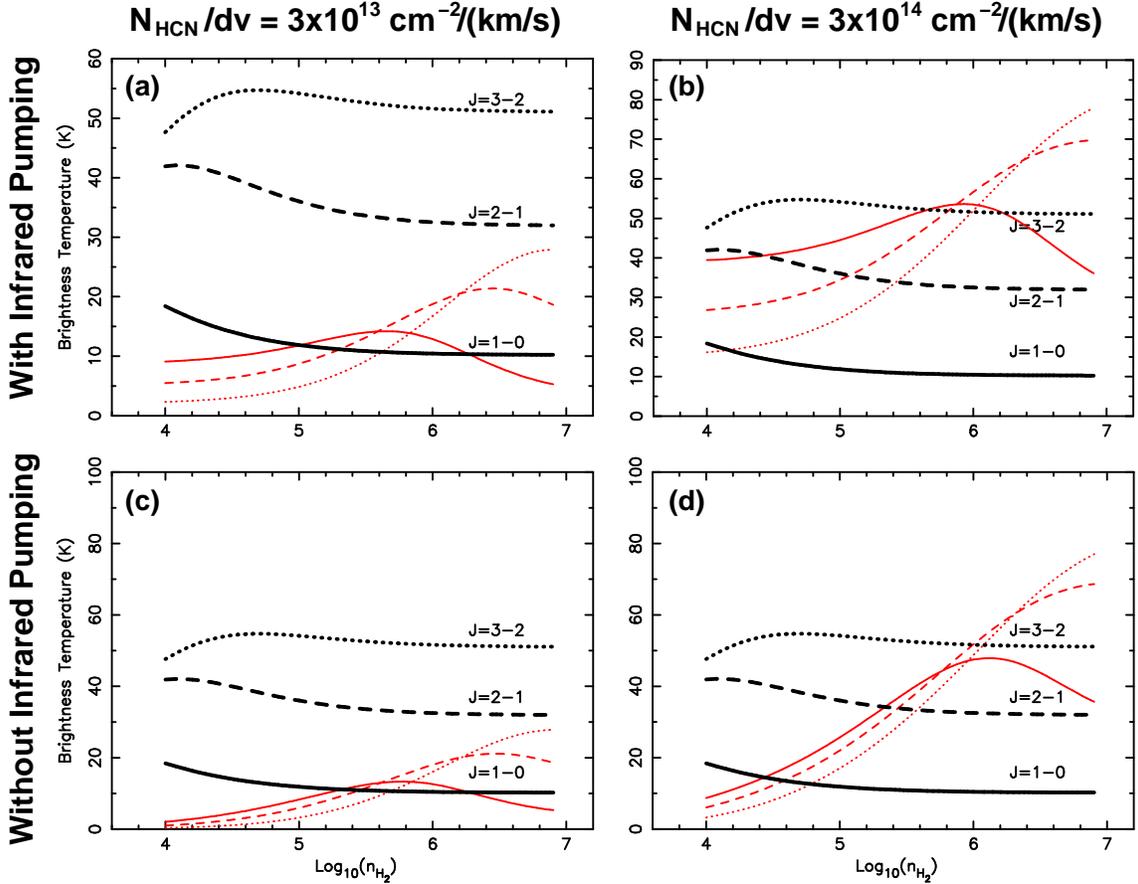}
\caption{LVG calculation results with (upper row plots) and without
	(bottom row plots) IR pumping.
	We adopted the kinetic temperature of molecular gas as 100 K.
	Plots are brightness temperatures (Kelvin) of CO (black lines)
	and HCN (red lines) as a function of H$_{2}$ number density
	(cm$^{-3}$).
	Solid, dashed, and dotted lines are the results for J = 1 -- 0,
	2 -- 1, and 3 -- 2 transitions, respectively.
	(a) Result with $N_{\rm HCN}/dv$ of $3\times10^{13}$ cm$^{-2}$
		(km s$^{-1})^{-1}$ and with IR pumping.
	(b) Result with $N_{\rm HCN}/dv$ of $3\times10^{14}$ cm$^{-2}$
		(km s$^{-1})^{-1}$ and with IR pumping.
	(c) Result with $N_{\rm HCN}/dv$ of $3\times10^{13}$ cm$^{-2}$
		(km s$^{-1})^{-1}$ but without IR pumping.
	(d) Result with $N_{\rm HCN}/dv$ of $3\times10^{14}$ cm$^{-2}$
		(km s$^{-1})^{-1}$ but without IR pumping.
\label{fig-IRP}}
\end{figure*}

If $f_{\rm HCN}$ is $\sim10$ times larger than $f_{\rm CO}$, then
it is possible to obtain the observed ratio, but it is physically
unlikely; the critical density of the HCN molecule is more than
an order of magnitude larger than that of the CO molecule, so that
wherever the HCN molecules can emit, the CO molecules can also emit,
and $f_{\rm HCN}$ cannot be larger than $f_{\rm CO}$.
A simple way to reproduce the observed HCN/CO ratio would be if the
[HCN]/[CO] abundance ratio is increased by a factor of $\sim10$.
In either case, the CO emission should be optically thin.
This is also true even for the non-LTE radiation transfer modeling
using the large-velocity-gradient (LVG) approximation;
without changing the HCN abundance, the HCN/CO ratio can be at most
around unity under very limited temperature or density range (see
Sect.~\ref{sect-dis-irpump} below and Fig.~\ref{fig-IRP}c).

\subsubsection{Excitation Mechanisms of the HCN Emission:
	II. Infrared Pumping}
\label{sect-dis-irpump}

In addition to the collisional excitation, here we also consider the
possibility of IR radiative pumping
\citep{mor75,car81,deg84}, since this mechanism can enhance the HCN
intensity as mentioned for the case of IRC$+10216$ in
Sect.~\ref{sect-dis-S2}.
This mechanism is to pump the population of a rotational level to
higher levels via vibrational transitions; in a strong IR radiation
field, some molecules that have vibrational transitions at IR
wavelength will be pumped to higher {\it vibrational} levels and
radiate back to lower levels, but to higher {\it rotational} levels
than before.
This process increases the population at higher rotational levels and
therefore makes the rotational transition emission stronger than
general molecular clouds (no strong IR field), which only have
collisional pumping.
This is particularly relevant when a luminous source is present
(e.g., AGN) in the background, such that optical/UV radiation is
absorbed by dust and reradiated in IR.
The vibrational transition of HCN molecule is at $14~\mu$m, where the
radiation from warm dust irradiated by AGNs is strong, and therefore
this is one of the possible mechanisms to explain the high HCN/CO
ratio toward AGNs.
The absorption line of the HCN $14~\mu$m transition can be seen in
several IR luminous galaxies \citep{lah07,far07}.
Recent interferometric observations toward those galaxies have indeed
displayed vibrationally excited HCN lines toward their IR luminous
nuclei \citep{sal08,sak10,ima13}.

To check whether IR pumping of HCN can occur near the nucleus, we
need to estimate the radiative excitation in this region.
The condition for the IR pumping to be effective is when the
probability of the HCN $v = 0 - 1$ vibrational transition absorption
at $14~\mu$m, $B_{v=0-1}I_{14\mu{\rm m}}$ (s$^{-1}$), is larger than
that of the HCN J = 1 -- 0 rotational transition spontaneous
emission, $A_{J=1-0}$ (s$^{-1}$), namely,
\begin{equation}
\label{eq-IRpump-condB}
B_{v=0-1}I_{14\mu{\rm m}} \ge A_{J=1-0},
\end{equation}
where $I_{14\mu{\rm m}}$ is the average intensity of the radiation
field at $14~\mu$m.
This equation can be re-written as
\begin{equation}
\label{eq-IRpump-condA}
\frac{c^{2}}{2h\nu^{3}}A_{v=1-0}I_{14\mu{\rm m}} \ge A_{J=1-0},
\end{equation}
where $c$ is the speed of light, $h$ is the Planck constant, and
$\nu$ is the frequency.
We use $A_{v=1-0}=2.23$ s$^{-1}$ \citep{har06} from the GEISA
spectroscopic database \citep{jac11} and
$A_{J=1-0}=2.41\times10^{-5}$ s$^{-1}$ \citep{dum10} from RADEX
\citep{tak07} for this calculation.

To estimate the radiation field at $14~\mu$m around the nucleus of
M51, we use the recent high resolution images taken with the
photometric bands S11 ($11~\mu$m) and L15 ($15~\mu$m) of the
{\it AKARI} Infrared Camera \citep{egu13}.
Both images show a point-like source at the nucleus of M51 at a
spatial resolution of $7\farcs4$, and well separated from the
surrounding spiral arms \citep[i.e., no contamination from the
emission of the spiral arms; see Fig.~2 of][]{egu13}.
We therefore assume that the point-like emission at the nucleus of
M51 in these images originates from the compact Seyfert 2 nucleus.
The flux densities of the nuclear emission at S11 and L15 are
$3.7\times10^{-2}$ Jy and $6.4\times10^{-2}$ Jy, respectively.
With linear interpolation between S11 and L15 flux densities, we
estimate a $14~\mu$m flux density, $S_{14\mu{\rm m}}$, of
$5.7\times10^{-2}$ Jy.

$I_{14\mu{\rm m}}$ can be calculated using $S_{14\mu{\rm m}}$ as
\begin{equation}
\label{eq-int}
I_{14\mu{\rm m}} = \frac{S_{14\mu{\rm m}}}{\Omega_{14\mu{\rm m}}},
\end{equation}
where $\Omega_{14\mu{\rm m}}$ is the solid angle of the source at
$14~\mu$m.
Since the compact infrared nuclear source of M51 has not been resolved
so far, we adopt the size of the infrared source as a few ($2-3$) pc,
which is the same as that of the thick dust disk observed toward the
nucleus of the prototypical Seyfert 2 galaxy NGC 1068 \citep{jaf04}.
Substituting the above values into eqs.~(\ref{eq-int}) and
(\ref{eq-IRpump-condA}), the left-hand side of the
eq.~(\ref{eq-IRpump-condA}) is calculated as
$(3.3-9.2)\times10^{-5}$ s$^{-1}$, which is larger than the
right-hand side of the equation of $2.41\times10^{-5}$ s$^{-1}$.
This indicates that IR pumping can occur in the molecular gas
surrounding the Seyfert 2 nucleus of M51.

We then included the infrared pumping effect in our LVG calculations.
We use the above calculated $14~\mu$m intensity for the radiation
coming from the background infrared source.
Here, we adopt the infrared source size of 3 pc.
We also assume that the molecular gas has a temperature of 100~K.
Note that in general higher molecular gas temperature induces more
collision, and therefore decreases the effect of infrared pumping.
But under the high temperature condition, the effect of infrared
pumping to lower transitions is insensitive to changes in the
molecular gas temperature, since most of population is at higher
transitions; a factor of a few change in temperature gives almost
no difference in the calculation results.

In the first calculation, we use the H$_{2}$ column density per
velocity of $1\times10^{21}$ cm$^{-2}$, which corresponds to the
H$_{2}$ column density toward the nucleus of M51
\citep[$6.2\times10^{21}$ cm$^{-2}$;][]{mat07} divided by the
typical velocity width of the extragalactic molecular clouds of
$6-8$ km s$^{-1}$ \citep{com97}.
We also assume that the column densities per velocity for CO and HCN,
$N_{\rm CO}/dv$ and $N_{\rm HCN}/dv$, are
$5\times10^{16}$ cm$^{-2}$ (km s$^{-1})^{-1}$ and
$3\times10^{13}$ cm$^{-2}$ (km s$^{-1})^{-1}$, respectively.
These column densities for CO and HCN correspond to the normal
abundance ratio mentioned in Sect.~\ref{sect-dis-lte} relative to the
H$_{2}$ column density mentioned above.
The results for the brightness temperatures of CO (black curves) and
HCN (red curves) lines are shown in Fig.~\ref{fig-IRP}(a), with the
comparison of the results without infrared pumping
(Fig.~\ref{fig-IRP}c).
As can be seen, using the normal abundance of HCN and CO, the
HCN(1 -- 0)/CO(1 -- 0) ratio of at most about unity can be realized
under very limited condition of $n_{\rm H_{2}}\sim10^{5-6}$
cm$^{-3}$, but cannot realize HCN(1 -- 0)/CO(1 -- 0) $>2$ both with
and without infrared pumping.

We then repeated the calculations by increasing the HCN abundance by
an order of magnitude, namely changing $N_{\rm HCN}/dv$ to
$3\times10^{14}$ cm$^{-2}$ (km s$^{-1})^{-1}$.
The calculation results are displayed in Fig.~\ref{fig-IRP}(b), with
the comparison of the results without infrared pumping
(Fig.~\ref{fig-IRP}d).
In both cases, high HCN(1 -- 0)/CO(1 -- 0) ratio of $>2$ can be
obtained at dense condition ($n_{\rm H_{2}}>10^{5}$~cm$^{-3}$).
On the other hand, when including infrared pumping,
HCN(1 -- 0)/CO(1 -- 0) $>2$ can also be obtained even at low density
condition, namely it is possible at large density range.

{\bf
Note that with the H$_{2}$ column density per velocity we used, the
CO brightness temperatures, especially J = 1 -- 0, are much lower
than the assumed molecular gas temperature of 100 K, since CO is
already thermalized and the population is at higher-J transitions,
which makes the lower-J lines optically thin and therefore low
brightness temperatures.
This is consistent with the non-detection of the CO lines toward
the nucleus under the current noise level of $\sim2-3$ K at $3\sigma$
upper limit (see Fig.~\ref{fig-ch480kms}) and the detection of the
HCN(1 -- 0) line.
}

In summary, infrared pumping does not need to be invoked to realize
high HCN(1 -- 0)/CO(1 -- 0) ratio of $>2$, but with infrared pumping,
it is easier to realize.
Since this is not conclusive only with the LVG calculations, we
discuss the possibility of infrared pumping with other observational
results in the next subsection.

\subsubsection{Possible Causes of High HCN Abundance}
\label{sect-dis-cause}

Based on the radiative transfer calculations above, it is necessary
to have high HCN abundance to realize high HCN(1 -- 0)/CO(1 -- 0)
ratio.
There are several ways in which the HCN abundance may be enhanced.
One possibility is due to high X-ray radiation from AGNs.
Since X-ray radiation penetrates much deeper and denser regions than
UV radiation, the chemical reactions exhibits different pathways
compared with Photo-Dissociated Region (PDR).
The molecular gas region that is irradiated by X-ray is called X-ray
Dominated Region \citep[XDR;][]{mal96}, and some radiative transfer
calculations of XDR succeeded to enhance the HCN abundance
\citep{lep96,lin06}, but some do not \citep{mei05,mei07}.
Second possibility is energetic particles (i.e., cosmic rays) either
from AGNs or radio jets.
This possibility, however, reduces the HCN and HNC abundances, and
increases the HCO$^{+}$ abundance \citep{bay10}, which do not match
with our results.
Another possibility is the high temperature gas phase chemistry as
recently suggested by \citet{har10,har13}.
This radiative transfer calculation includes the time-dependent
interstellar chemical gas-phase reaction network, so that this
calculation gives resultant molecular gas abundances under strong
heating by X-ray irradiation.
The calculation results show high HCN abundance, which matches with
our observations.
Furthermore, the HNC abundance increase is an order of magnitude less
than that for the HCN abundance, and the HCO$^{+}$ abundance shows
decrement under this model, which again matches with our results of
non-detection for the HCO$^{+}$(1 -- 0) and HNC(1 -- 0) lines.
Indeed, the HCO$^{+}$ vibrational absorption line at $12~\mu$m has
not been detected toward the luminous infrared galaxies from which
the HCN vibrational absorption line at $14~\mu$m has been detected
\citep{ima10}.
Since the HCO$^{+}$ vibrational transition is located at a similar
wavelength and has a similar Einstein coefficient as that of the HCN
vibrational transition, non-detection of the HCO$^{+}$ absorption
line suggests that the HCN abundance has enhanced compared to that of
the HCO$^{+}$, consistent with this model.
In addition, high HCN/HNC abundance ratios have been observed toward
active star forming cores \citep{gol86,sch92}, and low ratios in
starless cores \citep{hir98} in our Galaxy, which is again consistent
with this model.
Our results therefore strongly support the model presented in
\citet{har10,har13}.

On the other hand, the radiative transfer calculation results in
Sect.~\ref{sect-dis-irpump} cannot realize HCN(1 -- 0)/CO(3 -- 2)
above unity even under the HCN abundance enhanced conditions;
in the CO(3 -- 2) map (Fig.~\ref{fig-ch480kms}), contours at the
level of $3-4\sigma$, namely $1.3-1.7$~K, are at the nucleus, but no
intensity peak has been observed as in the HCN map, if we consider
the beam size.
This suggests that HCN(1 -- 0) is stronger than CO(3 -- 2) at the
nucleus, which is inconsistent with the calculation results in
Fig.~\ref{fig-IRP}.
One possible mechanism to make the HCN(1 -- 0) emission strong is
weak HCN masing.
The infrared pumping can invert the population at the HCN(1 -- 0)
transition and possible to produce the maser emission at this
transition \citep{din00}.
The HCN(1 -- 0) masers have indeed been observed toward various
carbon stars \citep{izu87,izu95,olo93}.
Under this excitation condition, and if the HCN molecule is
distributed as an edge-on torus or disk that is similar to the water
maser disk in NGC 4258 \citep{miy95}, it is possible to create HCN
maser emission at the systemic velocity.
This matches well with the strong HCN(1 -- 0) nuclear emission only
at the systemic velocity channel (Figs.~\ref{fig-ch480kms} and
\ref{fig-spec}).
Similar explanation has been discussed for the strong spike in the
HNC(3 -- 2) emission for the western nucleus of Arp 220
\citep{aal09}, although our observation has not been detected any
HNC emission, probably due to the high temperature gas phase
chemistry in our case, as mentioned above.

Our results therefore suggest that there is a compact dense
molecular gas in front of the type 2 AGN in M51 that cannot be
resolved in our spatial resolution of $\sim1''$ ($\sim34$~pc).
This component can be the obscuring material for the type 2 nucleus
of M51; assume this component has a size of $\sim1$ pc (i.e., much
smaller than the spatial resolution of our observations) along the
line of sight and a density of $\sim10^{6}$ cm$^{-3}$ (i.e.,
consistent with the radiative transfer calculation results in
Sect.~\ref{sect-dis-irpump}), then the column density along the line
of sight will be calculated as $\sim3\times10^{24}$ cm$^{-2}$.
This value is consistent with the column density of
$\sim5.6\times10^{24}$ cm$^{-2}$ for the absorption material in front
of the nuclear X-ray source (i.e., AGN) of M51 \citep{fuk01}.
It is highly required to observe with HCN lines with $\ll1''$
resolution toward M51 to directly image the obscuring material in
front of the AGN, which may give us knowledge about the still-unknown
obscuring and feeding mechanisms of AGNs.

\section{Summary}
\label{sect-sum}

Our newly obtained $\sim1''$ ($\sim34$ pc) resolution HCN(1 -- 0)
image toward the Seyfert 2 nucleus of M51 (NGC 5194) revealed the
following results:
\begin{itemize}
\item The overall HCN(1 -- 0) distribution and kinematics are similar
	to those of the CO lines.
	Since it is known that the kinematics derived from the CO lines
	indicates the jet-entrained nature, this similarity indicates
	that the dense gas is also entrained by the jets and outflowing
	from the AGN.
\item HCN(1 -- 0)/CO(1 -- 0) brightness temperature ratio is stronly
	enhanced along the jets with the value around unity, and the
	value is similar to that of the shocked gas in the outflows from
	young stellar objects in our Galaxy, suggesting that the HCN
	enhancement along the jets are highly related to shock.
\item A strong HCN(1 -- 0) emission has been detected toward the
	nucleus only at the systemic velocity channel.
	No CO emission has been detected at this region and at this
	velocity, and therefore the ratio turned out to be more than 2.
	To realize this high ratio, it is suggested from our radiative
	transfer model that
	weak HCN maser triggered by the infrared pumping with HCN
	abundance enhancement is the possible mechanism.
	The most plausible configuration for emitting the HCN maser is
	a rotating edge-on dense molecular gas disk or torus around the
	AGN, similar to the water maser disks in other Seyfert 2
	galaxies.
	If this disk or torus has the size of $\sim1$ pc and the density
	of $\sim10^{6}$ cm$^{-2}$, then it is possible to explain the
	nuclear obscuration for this type 2 AGN in M51.
\item Cause of the high HCN abundance can be due to high
	temperature gas phase chemistry.
	Non-detection of the HCO$^{+}$(1 -- 0) and HNC(1 -- 0) lines in
	our observations also support this model.
\end{itemize}

\acknowledgements
We would like to thank the IRAM and SMA staff for supporting our
observations.
We also thank Dr.\ T.\ Umemoto for the permission to use the
molecular line spectra of L1157, and Dr.\ F.\ Egusa for providing us
the flux information of the nucleus of M51 from the $AKARI$ Infrared
Camera data.
We are also grateful to the anonymous referee for very helpful
comments and discussion.
This collaboration was partly supported by the French-Taiwanese
Hubert Curien Program, ORCHID.
SM is supported by the National Science Council (NSC) and the
Ministry of Science and Technology (MoST) of Taiwan,
NSC 97-2112-M-001-021-MY3, NSC 100-2112-M-001-006-MY3, and
MoST 103-2112-M-001-032-MY3.
DVT acknowledges the financial support from National Foundation for
Science and Technology development (NAFOSTED) under the contract
103.99-2014.82.
DVT is funded by the project ``Development of problem-solving
environments for computational science based on resource sharing
high-performance computers at Vietnam Academy of Science and
Technology''.


\begin{thebibliography}{99}
\bibitem[Aalto et al.(1995)]{aal95} Aalto, S., Booth, R. S.,
	Black, J. H., \& Johansson, L. E. B.  1995, \aap, 300, 369
\bibitem[Aalto et al.(2012)]{aal12} Aalto, S., Garcia-Burillo, S.,
	Muller, S., Winters, J. M., van der Werf, P., Henkel, C.,
	Costagliola, F., \& Neri, R.  2012, \aap, 537, A44
\bibitem[Aalto et al.(2009)]{aal09} Aalto, S., Wilner, D.,
	Spaans, M., Wiedner, M. C., Sakamoto, K., Black, J. H.,
	\& Caldas, M.  2009, \aap, 493, 481
\bibitem[Antonucci(1993)]{ant93} Antonucci, R. R. J.  1993, \araa,
	31, 473
\bibitem[Bayet et al.(2010)]{bay10} Bayet, E., Hartquist, T. W.,
	Viti, S., Williams, D. A., \& Bell, T. A.  2010, \aap, 521, A16
\bibitem[Bradley et al.(2004)]{bra04} Bradley, L. D., Kaiser, M. E.,
	\& Baan, W. A.  2004, \apj, 603, 463
\bibitem[Brouillet et al.(2005)]{bro05} Brouillet, N., Muller, S.,
	Herpin, F., Braine, J., \& Jacq, T.  2005, \aap, 429, 153
\bibitem[Carrol \& Goldsmith(1981)]{car81} Carrol, T. J.,
	\& Goldsmith, P. F.  1981, \apj, 245, 891
\bibitem[Combes \& Becquaert(1997)]{com97} Combes, F.,
	\& Becquaert, J.-F.  1997, \aap, 326, 554
\bibitem[Combes et al.(2013)]{com13} Combes, F.,
	Garc\'ia-Burillo, S., Casasola, V., et al.  2013, \aap, 558, A124
\bibitem[Crane \& van der Hulst(1992)]{cra92} Crane, P. C.,
	\& van der Hulst, J. M.  1992, \aj, 103, 1146
\bibitem[Davies et al.(2009)]{dav09} Davies, R. I., Maciejewski, W.,
	Hicks, E. K. S., Tacconi, L. J., Genzel, R., \& Engel, H.  2009,
	\apj, 702, 114
\bibitem[Deguchi \& Uyemura(1984)]{deg84} Deguchi, S.,
	\& Uyemura, M.  1984, \apj, 285, 153
\bibitem[Dinh-V-Trung \& Nguyen-Q-Rieu(2000)]{din00} Dinh-V-Trung,
	\& Nguyen-Q-Rieu  2000, \aap, 361, 601
\bibitem[Dumouchel et al.(2010)]{dum10} Dumouchel, F., Faure, A.,
	\& Lique, F.  2010, \mnras, 406, 2488
\bibitem[Egusa et al.(2013)]{egu13} Egusa, F., Wada, T., Sakon, I.,
	Onaka, T., Arimatsu, K., \& Matsuhara, H.  2013, \apj, 778, 1
\bibitem[Farrah et al.(2007)]{far07} Farrah, D., Bernard-Salas, J.,
	Spoon, H. W. W., et al.  2007, \apj, 667, 149
\bibitem[Ford et al.(1985)]{for85} Ford, H. C., Crane, P. C.,
	Jacoby, G. H., Lawrie, D. G., \& van der Hulst, J. M.  1985,
	\apj, 293, 132
\bibitem[Fukazawa et al.(2001)]{fuk01} Fukazawa, Y., Iyomoto, N.,
	Kubota, A., Matsumoto, Y., \& Makishima, K.  2001, \aap, 374, 73
\bibitem[Gao \& Solomon(2004)]{gao04} Gao, Y.,
	\& Solomon, P. M.  2004, \apj, 606, 271
\bibitem[Goldsmith et al.(1986)]{gol86} Goldsmith, P. F.,
	Irvine, W. M., Hjalmarson, A., \& Ellder, J.  1986,
	\apj, 310, 383
\bibitem[Hagiwara et al.(2001)]{hag01} Hagiwara, Y., Henkel, C.,
	Menten, K. M., \& Nakai, N.  2001, \apjl, 560, L37
\bibitem[Harada et al.(2010)]{har10} Harada, N., Herbst, E.,
	\& Wakelam, V.  2010, \apj, 721, 1570
\bibitem[Harada et al.(2013)]{har13} Harada, N., Thompson, T. A.,
	\& Herbst, E.  2013, \apj, 765, 108
\bibitem[Harris et al.(2006)]{har06} Harris, G. J., Tennyson, J.,
	Kaminsky, B. M., Pavlenko, Ya. V., \& Jones, H. R. A.  2006,
	\mnras, 367, 400
\bibitem[Helfer \& Blitz(1993)]{hel93} Helfer, T. T.,
	\& Blitz, L.  1993, \apj, 419, 86
\bibitem[Helfer \& Blitz(1997)]{hel97} Helfer, T. T.,
	\& Blitz, L.  1997, \apj, 478, 233
\bibitem[Hicks et al.(2009)]{hic09} Hicks, E. K. S., Davies, R. I.,
	Malkan, M. A., Genzel, R., Tacconi, L. J.,
	M\"uller S\'anchez, F., \& Sternberg, A.  2009, \apj, 696, 448
\bibitem[Hirota et al.(1998)]{hir98} Hirota, T., Yamamoto, S.,
	Mikami, H., \& Ohishi, M.  1998, \apj, 503, 717
\bibitem[Ho et al.(1997)]{ho97} Ho, L. C., Filippenko, A. V.,
	\& Sargent, W. L. W.  1997, \apjs, 112, 315
\bibitem[Ho et al.(2004)]{ho04} Ho, P. T. P., Moran, J. M.,
	\& Lo, F.  2004, \apj, 616, L1
\bibitem[Hsieh et al.(2008)]{hsi08} Hsieh, P.-Y., Matsushita. S.,
	Lim, J., Kohno, K., \& Sawada-Satoh, S.  2008, \apj, 683, 70
\bibitem[Imanishi et al.(2013)]{ima13} Imanishi, M.,
	\& Nakanishi, K.  2013, \aj, 146, 91
\bibitem[Imanishi et al.(2010)]{ima10} Imanishi, M., Nakanishi, K.,
	Yamada, M., Tamura, Y., \& Kohno, K.  2010, \pasj, 62, 201
\bibitem[Irvine et al.(1987)]{irv87} Irvine, W. M., Goldsmith, P.F.,
	\& Hjalmarson, A.  1987, in Interstellar Process,
	ed. D. J. Hollenbach \& H. A. Thronson (Dordrecht: Reidel), 585
\bibitem[Izumiura et al.(1987)]{izu87} Izumiura, H., Ukita, N.,
	Kawabe, R., Kaifu, N., Tsuji, T., Unno, W., \& Koyama, K. 1987,
	\apjl, 323, L81
\bibitem[Izumiura et al.(1995)]{izu95} Izumiura, H., Ukita, N.,
	\& Tsuji, T.  1995, \apj, 440, 728
\bibitem[Jackson et al.(1996)]{jac96} Jackson, J. M., Heyer, M. H.,
	Paglione, T. A. D., \& Bolatto, A. D.  1996, \apjl, 456, L91
\bibitem[Jackson et al.(1993)]{jac93} Jackson, J. M.,
	Paglione, T. A. D., Ishizuki, S., \& Nguyen-Q-Rieu  1993,
	\apjl, 418, L13
\bibitem[Jacquinet-Husson et al.(2011)]{jac11} Jacquinet-Husson, N.,
	Crepeau, L., Armante, R., et al.  2011, \jqsrt, 112, 2395
\bibitem[Jaffe et al.(2004)]{jaf04} Jaffe, W., Meisenheimer, K.,
	\& R\"ottgering, H. J. A.  2004, \nat, 429, 47
\bibitem[Kohno et al.(1996)]{koh96} Kohno, K., Kawabe, R., Tosaki T.,
	\& Okumura S. K.  1996, \apjl, 461, L29
\bibitem[Kohno et al.(2001)]{koh01} Kohno, K., Matsushita, S.,
	Vila-Vilar\'o, B., et al.  2001, ASP Conf.\ Ser.\ 249, 672
\bibitem[Kohno et al.(2008a)]{koh08a} Kohno, K., Muraoka, K..
	Hatsukade, B., et al.  2008, in EAS Publ.\ Ser.\ 31, 65
\bibitem[Kohno et al.(2008b)]{koh08b} Kohno, K., Nakanishi, K.,
	Tosaki, T., Muraoka, K., Miura, R., Ezawa, H.,
	\& Kawabe, R.  2008, \apss, 313, 279
\bibitem[Krips et al.(2011)]{kri11} Krips, M., Mart\'in, S.,
	Eckart, A., et al.  2011 \apj, 736, 37
\bibitem[Krips et al.(2007)]{kri07} Krips, M., Neri, R.,
	Garc\'ia-Burillo, S., et al.  2007, \aap, 468, L63
\bibitem[Lahuis et al.(2007)]{lah07} Lahuis, F., Spoon, H. W. W.,
	Tielens, A. G. G. M., et al.  2007, \apj, 659, 296
\bibitem[Lepp \& Dargarno(1996)]{lep96} Lepp, S.,
	\& Dalgarno, A.  1996, \aap, 306, L21
\bibitem[Lintott \& Viti(2006)]{lin06} Lintott, C.,
	\& Viti, S.  2006, \apjl, 646, L37
\bibitem[Maloney et al.(1996)]{mal96} Maloney, P. R.,
	Hollenbach, D. J., \& Tielens, A. G. G. M.  1996, \apj, 466, 561
\bibitem[Matsushita et al.(2010)]{mat10} Matsushita, S., Kawabe, R.,
	Kohno, K., Tosaki, T., \& Vila-Vilar\'o, B.  2010, \pasj, 62, 409
\bibitem[Matsushita et al.(1998)]{mat98} Matsushita, S., Kohno, K.,
	Vila-Vilar\'o, B., Tosaki, T., \& Kawabe, R., 1998,
	\apj, 495, 267
\bibitem[Matsushita et al.(2007)]{mat07} Matsushita, S., Muller, S.,
	\& Lim, J.  2007, \aap, 468, L49
\bibitem[Matsushita et al.(2004)]{mat04} Matsushita, S., Sakamoto,
	K., Kuo, C.-Y., et al.  2004, \apjl, 616, L55
\bibitem[Meijerink \& Spaans(2005)]{mei05} Meijerink, R.,
	\& Spaans, M.  2007, \aap, 436, 397
\bibitem[Meijerink et al.(2007)]{mei07} Meijerink, R., Spaans, M.,
	\& Israel, F. P.  2007, \aap, 461, 793
\bibitem[Miyoshi et al.(1995)]{miy95} Miyoshi, M., Moran, J.,
	Herrnstein, J., Greenhill, L., Nakai, N., Diamond, P.,
	\& Inoue, M.  1995, \nat, 373, 127
\bibitem[Morris(1975)]{mor75} Morris, M.  1975, \apj, 197, 603
\bibitem[M\"uller S\'anchez et al.(2009)]{mul09}
	M\"uller S\'anchez, F., Davies, R. I., Genzel, R.,
	Tacconi, L. J., Eisenhauer, F., Hicks, E. K. S., Friedrich, S.,
	\& Sternberg, A.  2009, \apj, 691, 749
\bibitem[Olofsson et al.(1993)]{olo93} Olofsson, H., Eriksson, K.,
	Gustafsson, B., \& Carlstrom, U.  1993, \apjs, 87, 305
\bibitem[Sakamoto et al.(2010)]{sak10} Sakamoto, K., Aalto, S.,
	Evans, A. S., Wiedner, M. C., \& Wilner, D. J.  2010,
	\apjl, 725, L228
\bibitem[Salter et al.(2008)]{sal08} Salter, C. J., Ghosh, T.,
	Catinella, B., et al.  2008, \aj, 136, 389
\bibitem[Sani et al.(2012)]{san12} Sani, E., Davies, R. I.,
	Sternberg, A., et al.  2012, \mnras, 424, 1963
\bibitem[Schilke et al.(1992)]{sch92} Schilke, P., Walmsley, C. M.,
	Pineau Des Forets, G., Roueff, E., Flower, D. R.,
	\& Guilloteau, S.  1992, \aap, 256, 595
\bibitem[Scoville et al.(1998)]{sco98} Scoville, N. Z., Yun, M. S.,
	Armus, L., \& Ford, H.  1998, \apjl, 493, L63
\bibitem[Shetty et al.(2007)]{she07} Shetty, R., Vogel, S. N.,
	Ostriker, E. C., \& Teuben, P. J.  2007, \apj, 665, 1138
\bibitem[Solomon et al.(1992)]{sol92} Solomon, P. M., Downes, D.,
	\& Radford, S. J. E.  1992, \apjl, 387, L55
\bibitem[Sorai et al.(2002)]{sor02} Sorai, K., Nakai, N., Kuno, N.,
	\& Nishiyama, K.  2002, \pasj, 54, 179
\bibitem[Tak\'ats \& Vink\'o(2006)]{tak06} Tak\'ats, K.,
	\& Vink\'o, J.  2006, \mnras, 372, 1735
\bibitem[Tully(1988)]{tul88} Tully, R. B.  1988,
	Nearby Galaxies Catalog (Cambridge: Cambridge Univ. Press)
\bibitem[Umemoto et al.(1992)]{ume92} Umemoto, T., Iwata, T.,
	Fukui, Y., Mikami, H., Yamamoto, S., Kameya, O.,
	\& Hirano, N.  1992, \apjl, 392, L83
\bibitem[Van der Tak et al.(2007)]{tak07}  Van der Tak, F. F. S.,
	Black, J. H., Sch\"oier, F. L., Jansen, D. J.,
	\& van Dishoeck, E. F.  2007, \aap, 468, 627
\bibitem[Wright et al.(1996)]{wri96} Wright, M. C. H.,
	Plambeck, R. L., \& Wilner, D. J.  1996, \apj, 469, 216
\end{thebibliography}
\end{document}